\documentclass[aps,prl,showpacs,twocolumn]{revtex4}
\usepackage{graphics} 
\usepackage{amssymb}
\usepackage{amsmath}
\usepackage{bm}

\usepackage[pdftex]{graphicx}
\pdfoutput=1
   
\begin{document}

\title{Observation of Thermally Activated Vortex Pairs in a Quasi-2D Bose Gas}

\author{Jae-yoon Choi}
\author{Sang Won Seo}
\author{Yong-il Shin}\email{yishin@snu.ac.kr}

\affiliation{Center for Subwavelength Optics and Department of Physics and Astronomy, Seoul National University, Seoul 151-747, Korea}

\date{\today}

\begin{abstract}
We measure the in-plane distribution of thermally activated vortices in a trapped quasi-2D Bose gas, where we enhance the visibility of density-depleted vortex cores by radially compressing the sample before releasing the trap. The pairing of vortices is revealed by the two-vortex spatial correlation function obtained from the vortex distribution. The vortex density decreases gradually as temperature is lowered, and below a certain temperature, a vortex-free region emerges in the center of the sample. This shows the crossover from a Berezinskii-Kosterlitz-Thouless phase containing vortex-pair excitations to a vortex-free Bose-Einstein condensate in a finite-size 2D system.
\end{abstract}


\maketitle
 
Understanding the emerging mechanisms of superfluidity has been a central theme in many-body physics. In particular, the superfluid state in a two-dimensional (2D) system is intriguing because formation of long-range order is prohibited by large thermal fluctuations~\cite{Mermin_PRL,Hohenber_PR} and consequently the picture of Bose-Einstein condensation is not applicable to the phase transition. The Berezinskii-Kosterlitz-Thouless (BKT) theory provides a microscopic mechanism for the 2D phase transition~\cite{B_JETP,KT_JP}, where vortices with opposite circulation are paired below a critical temperature. Because vortex-antivortex pairs carry a zero net phase slip on the large length scale compared to the vortex pair size, the decay of the phase coherence changes from exponential to algebraic. The BKT mechanism has been experimentally tested in many 2D systems~\cite{He_PRL,JJA_PRL,quasi_H,BKTcross_Nature,BKTmech_PRL}.

Ultracold atomic gases in 2D geometry present a clean and well-controlled system for studying BKT physics. Previous experimental studies of phase coherence~\cite{BKTcross_Nature,2DNIST_PRL} and thermodynamic properties~\cite{PreSF_PRL,Scale_Nat,LDA_PRL} showed that a BKT-type transition occurs in a quasi-2D Bose gas trapped in a harmonic potential. Recently, superfluid behavior was demonstrated by measuring a critical velocity for friction-less motion of an obstacle~\cite{SF2d_Natphys}. One of the appealing features of the ultracold atomic gas system is that individual vortices can be detected, which would provide unique opportunities to investigate the details of the microscopic nature of the BKT transition. The pairing of vortices is the essential part for establishing quasi-long-range order in the 2D superfluid. However, its direct observation has been elusive so far.

In this Letter, we report the observation of thermally activated vortex pairs in a trapped quasi-2D Bose gas. We measure the in-plane vortex distribution of the sample by detecting density-depleted vortex cores, and observe that the two-vortex spatial correlation function obtained from the vortex distribution shows the pairing of vortices. We investigate the temperature evolution of the vortex distribution. As the temperature is lowered, the vortex density decreases, preferentially in the center of the sample, and eventually, a vortex-free region emerges below a certain temperature. This manifests the crossover of the superfluid from a BKT phase containing thermally activated vortices to a Bose-Einstein condensate (BEC). In a finite-size 2D system, forming a BEC is expected at finite temperature when the coherence length becomes comparable to the spatial extent of the system~\cite{BEC-BKTtransition,Flatland_Natview}. Our results clarify the nature of the superfluid state in an interacting 2D Bose gas trapped in a harmonic potential.

\begin{figure}
\includegraphics[width=8.5cm]{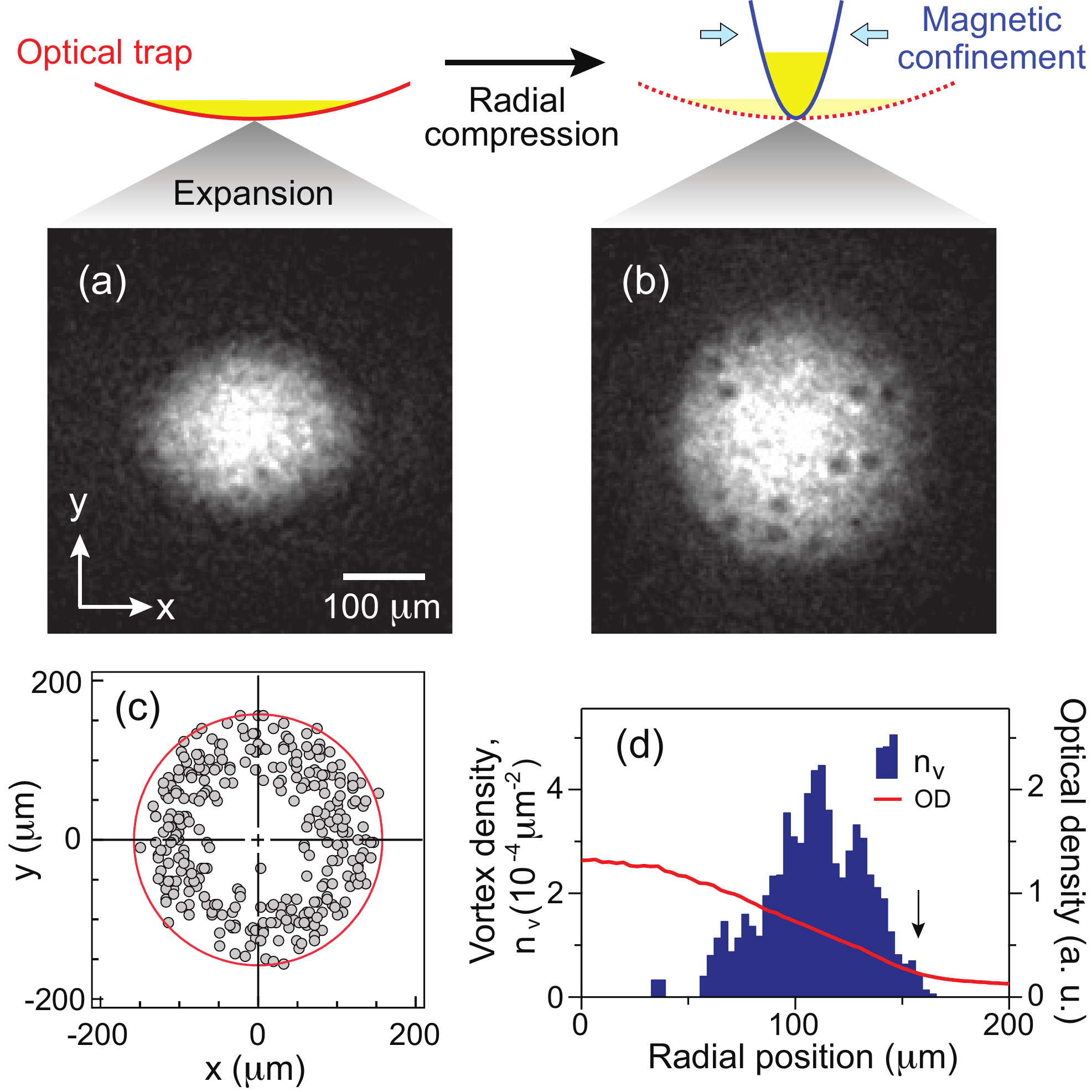}
\caption{(color online). Observation of thermally activated vortices. (a) Optical density image of a quasi-2D Bose gas, taken after a 17-ms time-of-flight expansion. Density ripples develop during the expansion due to phase fluctuations in the sample. (b) Image with applying a radial compression before expansion. The radial compression is achieved by superposing a magnetic potential onto the sample in an optical trap, and the expansion includes a 6-ms expansion in the optical trap and a subsequent 9-ms time-of-flight. Quantized vortices are observed with density-depleted cores. (c) Positions of vortices with respect to the sample center, recorded for 20 realizations of the same experiment. (d) Radial profile of the vortex density $n_v(r)$. The red solid line displays the atom density profile $n(r)$ of the sample in (b), showing a bimodal distribution. The center part is refered to as the coheret part of the sample and its boundary $R_c$ is indicated by a red solid line in (c) and an arrow in (d). The data are acquired for $N=1.6(2)\times10^6$ atoms at $T=50(6)$~nK. The vortex number in an image is $N_v=16(4)$. }
\label{Figure1}
\end{figure}

Our experiments are carried out with a quasi-2D Bose gas of $^{23}$Na atoms in a pancake-shaped optical dipole trap~\cite{phasefluc_PRL}. The trapping frequencies of the harmonic potential are $(\omega_x, \omega_y, \omega_z)=2\pi\times (3.0, 3.9, 370)$ Hz, where the $z$-axis is along the gravity direction. The dimensionless interaction strength is $\tilde{g}= \sqrt{8\pi}a/l_z\simeq0.013$, where $a$ is the three-dimensional (3D) scattering length and $l_z=\sqrt{\hbar/m\omega_z}$ is the axial harmonic oscillator length ($m$ is the atomic mass). For a typical sample of $N=1.3\times10^6$ atoms, the zero-temperature chemical potential is estimated to be $\mu\simeq0.7\hbar \omega_z$, where $\hbar$ is the Planck constant $h$ divided by $2\pi$. For our small $\tilde{g}$, the BKT critical temperature is estimated to be $T_c\approx 0.99 T_0 =75$~nK in the mean-field theory~\cite{Holzmann_EPL}, where $T_0$ is the Bose-Einstein condensation temperature for a quasi-2D, non-interacting ideal gas of $N$ atoms in the harmonic trap~\cite{footnote1}. Below the critical temperature, the sample shows a bimodal density distribution after time-of-flight expansion~\cite{Bimodal2D_PRL}. We refer to the center part as the coherent part of the sample and determine the sample temperature $T$ from a gaussian fit to the outer thermal wings.

The conventional method for detecting quantized vortices is observing density-depleted vortex cores after releasing the trap~\cite{Vimage_PRL}. Because the vortex core size $\xi\propto n^{-1/2}\propto R^{3/2}$, where $n$ is the atom density and $R$ is the sample radius, the vortex core expands faster than the sample in a typical 3D case~\cite{Dalfovo_PRA}, facilitating its detection. However, this simple method is not adequate for detecting thermally activated vortices in a 2D Bose gas. The fast expansion of the sample along the tight direction rapidly reduces the atom interaction effects so that phase fluctuations due to thermal excitations of vortices as well as phonons evolve into complicated density ripples as a result of self-interference [Fig.~1(a)]~\cite{phasefluc_PRL}. Thus, it is impossible to unambiguously distinguish individual vortices in the image of the simply expanding 2D Bose gas.

In order to enhance the visibility of the vortex cores, we apply a radial compression to the sample before releasing the trap. This compression transforms the 2D sample into an oblate 3D sample with $\mu/\hbar\omega_z>5$, restoring the favorable condition for the vortex detection~\cite{Dalfovo_PRA}. Also, the density of states changes to 3D, inducing thermal relaxation of phonons. Although tightly bound vortex pairs might annihilate, loosely bound pairs and free vortices would survive the compression process~\cite{BKTmech_PRL}.

In our experiments, the radial compression is achieved by superposing a magnetic potential onto the sample~\cite{Skyrmion}. We increase the radial trapping frequencies at the center of the hybrid trap to $2\pi\times39$~Hz for 0.4~s without any significant collective oscillation of the sample, and let the sample relax for 0.2~s which corresponds to about 10 collision times~\cite{supplementary}. Finally, we probe the in-plane atom density distribution by absorption imaging after 15~ms expansion. We first turn off the magnetic potential and switch off the optical trap 6~ms later, which we find helpful to improve the core visibility in our experiments. 


\begin{figure}
\includegraphics[width=7.3cm]{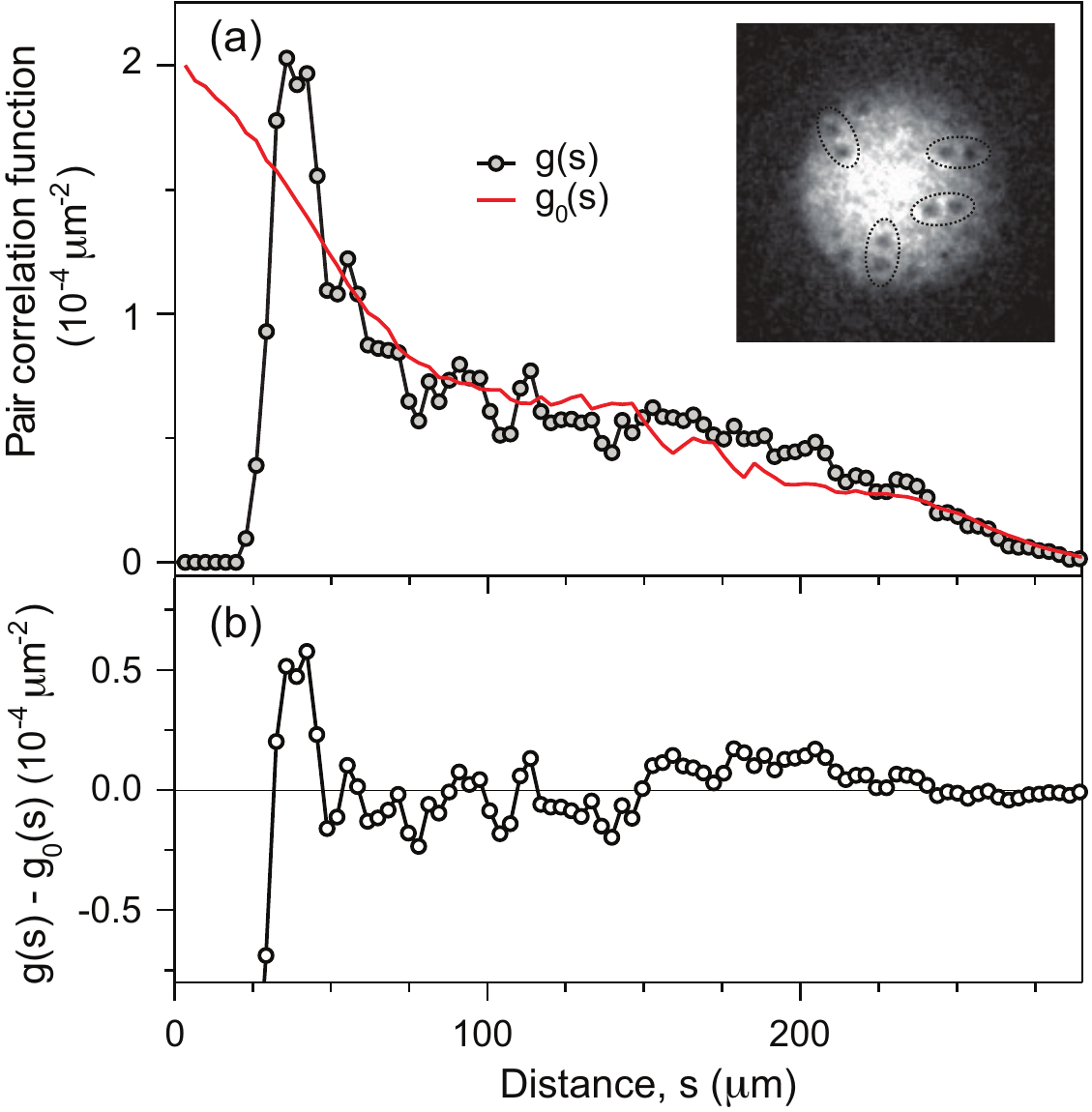}
\caption{(color online). Pair correlations of vortices. (a) Two-vortex spatial correlation function $g(s)$ for the experimental condition in Fig.~1. The red solid line represents the correlation function $g_0(s)$ for a random distribution with the same vortex density profile $n_v(r)$ in Fig.~1(d). (b) The difference $g(s)-g_0(s)$ shows oscillatory behavior with a noticeable enhancement at $s\sim35~\mu$m. The inset displays the same image in Fig.~1(b) with guide lines for pairs of vortices whose separation is less than 30~$\mu$m.}
\label{Figure2}
\end{figure}

We observe thermally activated vortices with clear density-depleted cores [Fig.~1(b)], where the visibility of the cores is about 50\% and the full width at half maximum depletion is $\xi_v\approx 8~\mu$m. By locating the vortex positions by hand, we obtain the vortex distribution $\rho_v(\vec{r})=\Sigma_{i=1}^{N_v} \delta(\vec{r}-\vec{r}_i)$, where $N_v$ is the number of vortices in the image and $\vec{r}_i$ denotes the vortex position with respect to the center of the sample.  The averaged vortex distribution $\bar{\rho}_v(\vec{r})$ shows no azimuthal dependence [Fig.~1(c)] and we obtain the radial profile of the vortex density $n_v(r)$ by azimuthally averaging $\bar{\rho}_v(\vec{r})$ [Fig.~1(d)]. Vortices mainly appear in the outer region of the coherent part, implying a vortex-driven phase transition. At the boundary of the coherent part, density modulations suggestive of vortex cores are often observed, but not included in the vortex counting if they have no local density-minimum points. 

When samples were prepared in a slightly tighter trap composed of the optical trap and a weak magnetic potential ($\mu/\hbar\omega_z\approx1.5$), we observed that the vortex number rapidly decreases in comparison to the samples prepared in the original optical trap at similar atom-number and temperature conditions ($\mu/\hbar\omega_z<1$)~\cite{supplementary}. This shows that the observed vortex excitations result from the 2D nature of the system.

One notable feature in the vortex distribution is frequent appearance of a pair of vortices which are closely located to each other but well separated from the others (the inset in Fig.~2). The average distance to the nearest-neighbor vortex is measured to be $d_m\approx 4\xi_v$. Since direct generation of single vortices is forbidden by the angular momentum conservation, the thermal activation of vortices should involve vortex-antivortex-pair excitations. Therefore, we infer that the observed vortex pairs consist of vortices with opposite circulation, attracting each other.

A more quantitative study of the pair correlations is performed with the two-vortex spatial correlation function
\begin{equation}
g(s)= \frac{1}{\pi{}sN_v(N_v-1)} \sum_{i>j} \delta(s-r_{ij})
\end{equation}
where $r_{ij}=|\vec{r}_i-\vec{r}_j|$. This function displays the probability of finding a vortex at distance $s$ from another vortex, reflecting the vortex-vortex interaction effects~\cite{VortexInteraction}. We determine $g(s)$ as the average of the pair correlation functions obtained from individual images for the same experiment [Fig.~2(a)]. In order to extract the pairing features, we compare $g(s)$ to the correlation function $g_0(s)$ for a random distribution with the same vortex density profile $n_v(r)$, which is calculated as $g_0(s)= \int drd\theta~r n_v(r) n_v(r')/ [\int dr 2\pi r n_v(r)]^2$ where $r'=\sqrt{r^2+2rs\cos\theta+s^2}$. The difference $g(s)-g_0(s)$ shows a noticeable enhancement around $s\sim d_m$ and small oscillatory behavior for $s>d_m$ [Fig.~2(b)], indicating the attraction between closely located vortices. The strong suppression at $s<3\xi_v$ is attributed to the annihilation of tightly bound vortex pairs during the compression process. These observations clearly demonstrate the pairing of vortices in the superfluid phase in a quasi-2D Bose gas.

\begin{figure}
\includegraphics[width=8.5cm]{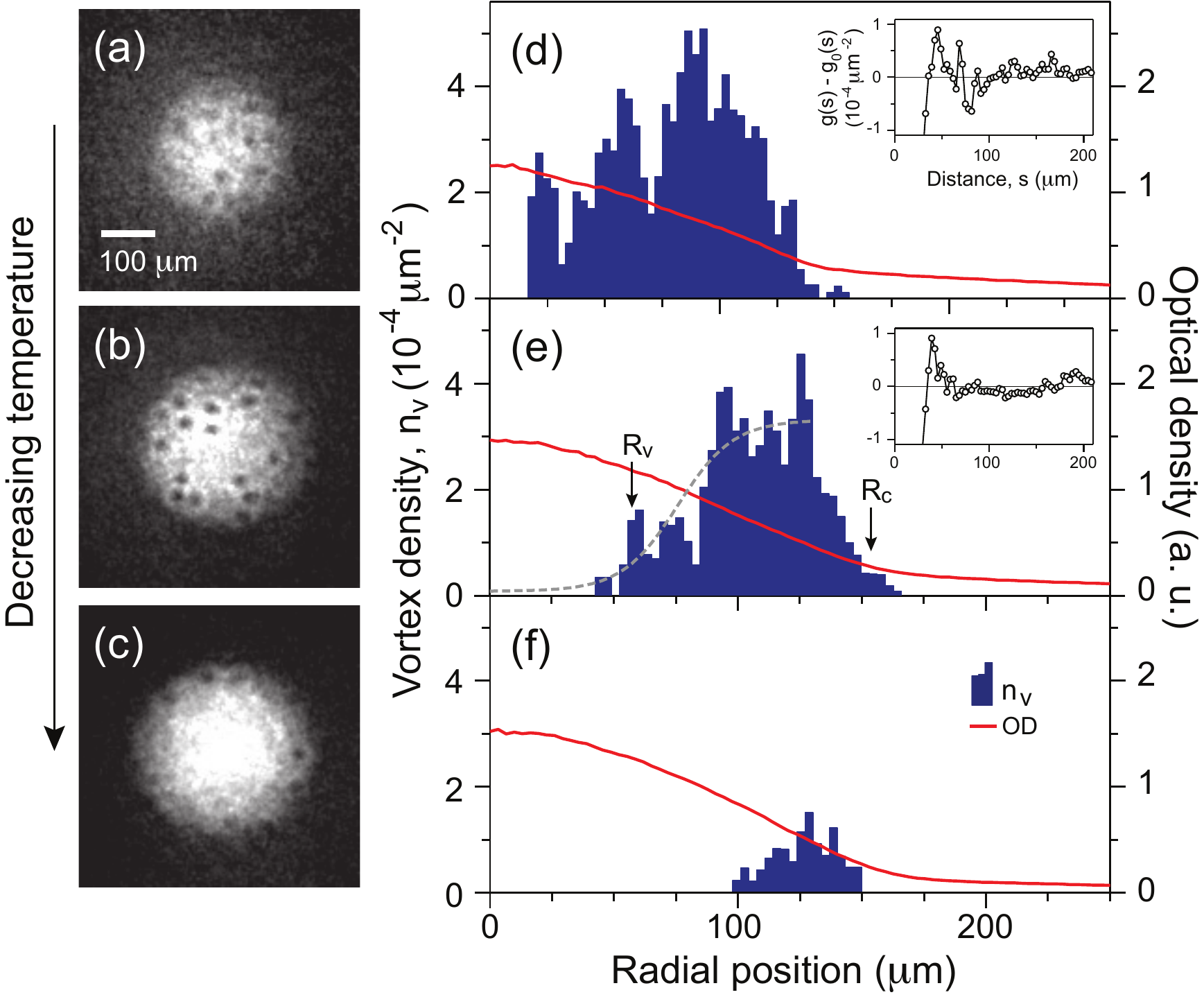}
\caption{(color online). Temperature dependence of the vortex distribution in a trapped 2D Bose gas. (a-c) Images of samples for various temperatures and (d-f) the corresponding vortex density profiles $n_v(r)$. The insets show the difference of the pair correlations functions, $g(s)-g_0(s)$. For the lowest temperature, $g(s)$ is not reliably determined due to the small vortex number. The inner boundary $R_v$ of the vortex region is determined from a hyperbolic-tangent fit (dashed) to the inner increasing part of $n_v(r)$. The sample conditions and the vortex numbers are $(N, T, N_v)=(1.5(2)\times 10^6, 67(12)$~nK, 13(4)) for (a,d), $(1.3(1)\times 10^6, 48(8)$~nK, 15(5)) for (b,e), and $(1.0(3)\times 10^6, 23(4)$~nK, 2(4)) for (c,f). The data are obtained from at least 12 images for each condition.}
\label{Figure3}
\end{figure}

We study the temperature evolution of the vortex distribution. Fig.~3 displays the vortex density profiles $n_v(r)$ for various temperatures. At high temperature below the critical point, vortex excitations prevail over the whole coherent part. As the temperature is lowered, the vortex excitations are suppressed preferentially in the center of the sample, and at the lowest temperature, only a few vortices appear near the boundary of the coherent part. The pairing features in the two-vortex spatial correlation function is preserved over all temperatures (the insets in Fig.~3).

The evolution of $n_v(r)$ can be qualitatively understood with estimating the thermal excitation probability $p$ of a vortex-antivortex pair in a uniform superfluid. The excitation energy of a vortex pair is $E\sim \frac{h^2}{2\pi m} n_s \ln (d/\xi)$ where $n_s$ is the superfluid density and $d$ is the separation of the vortices~\cite{BEC-BKTtransition}. In the superfluid of radius $R$, the number of distinguishable microstates for the vortex pair is $\sim \frac{R^2}{\xi^2}\frac{\pi d}{\xi}$, and the entropy $S=k \ln (\pi d R^2/\xi^3)$ ($k$ is the Boltzmann constant). The associated free energy $F=E-TS$ gives $p\propto e^{-F/kT}=\frac{\pi R^2}{\xi^2}(d/\xi)^{-(n_s \lambda^2-1)}$, which is exponentially suppressed by $n_s \lambda^2$, where $\lambda=\frac{h}{\sqrt{2\pi m kT}}$ is the thermal wavelength. Thus, the vortex density profile $n_v(r)$ reflects the spatial distribution of the superfluid density $n_s(r)$ in the inhomogeneous trapped sample. Distortion of the vortex distribution might be anticipated due to vortex diffusion during the compression time. Indeed, a vortex pair carries a linear momentum of $h/d$. However, we observe that the vortex region ($n_v>0$) does not change significantly for longer compression times, while the vortex density decreases, which is attributed to the vortex-pair annihilation~\cite{supplementary}.

\begin{figure}
\includegraphics[width=7.5cm]{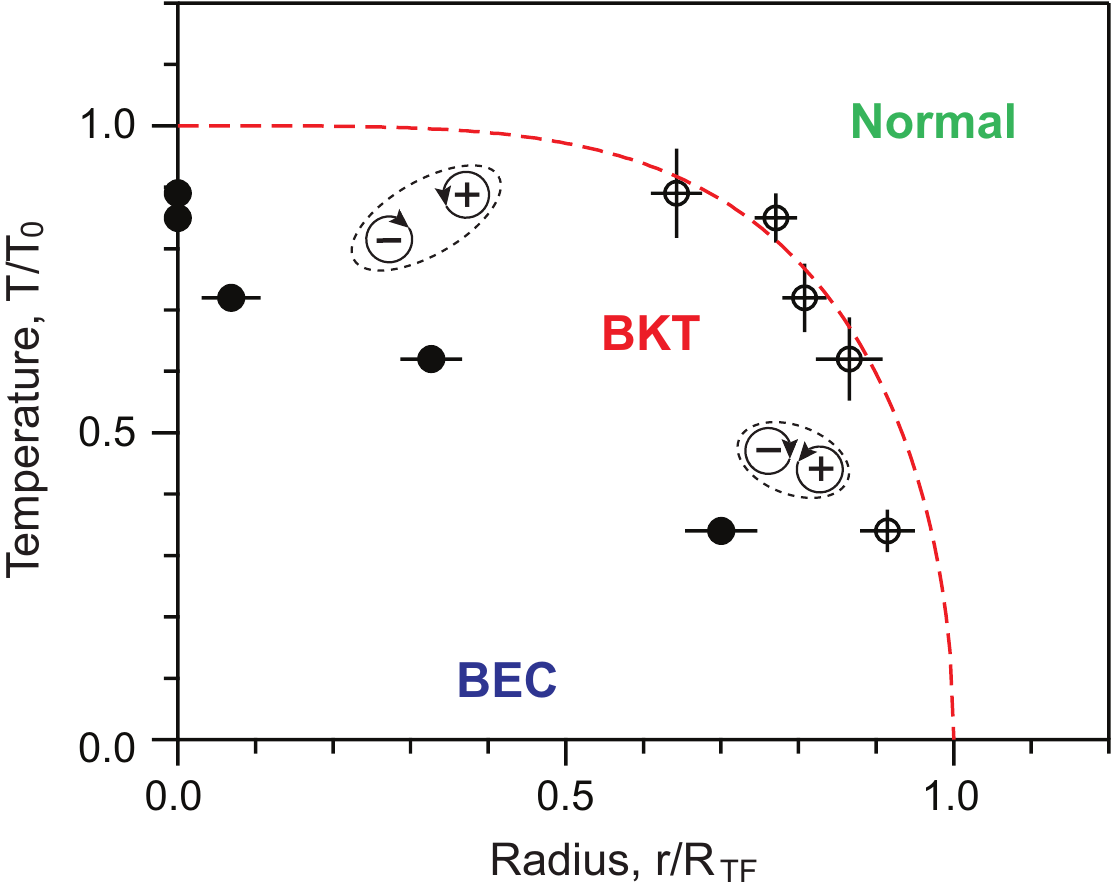}
\caption{(color online). BKT-BEC crossover. A schematic phase diagram of a weakly interacting 2D Bose gas trapped in a harmonic potential is depicted in the plane of $T/T_0$ and $r/R_\textrm{TF}$. $T_0$ is the BEC temperature for a quasi-2D ideal gas~\cite{footnote1}, and $R_\textrm{TF}$ is the zero-temperature 2D Thomas-Fermi radius. Open and solid circles indicate the radii of the coherent part and the inner radii of the vortex region ($n_v>0$), respectively~\cite{supplementary}. The dashed line is a guide for the critical line for the superfluid transition. As thermal excitations of vortex pairs are suppressed at lower temperatures, the BKT superfluid evolves into a BEC in the finte system.}
\label{Figure4}
\end{figure}

In a trapped 2D system, the density of states is modified by the trapping potential so that a BEC can form even at finite temperature~\cite{2DBEC}. This suggests that the superfluid in the trapped 2D Bose gas would evolve into a BEC at low termperatures, where the phase coherence becomes extended over the whole superfluid via suppressing thermal excitations of vortices~\cite{BEC-BKTtransition,Flatland_Natview}. We observe that a vortex-free region emerges below a certain temperature, and this is a manifestation of the BKT-BEC crossover behavior of the system. We determine the inner boundary $R_v$ of the vortex region from a hyperbolic-tangent fit to the inner increasing part of $n_v(r)$ with a threshold value of $6\times10^{-5}/\mu$m$^2$ [Fig.~3(e)], which defines the characteristic radius for the BKT-BEC crossover. 

We summarize our results in Fig.~4 with a schematic phase diagram for a trapped 2D Bose gas in the plane of temperature and radial position. Here we use the coherent part as a marker for the superfluid phase transition~\cite{Bimodal2D_PRL}. The in-situ radius $R_c$ is determined from a bimodal fit to the density profile in the images taken without the compression [Fig.~1(a)], including the time-of-flight expansion factor, and $R_v/R_c$ is measured from the images taken with the compression [Fig.~1(b)]~\cite{supplementary}. In recent theoretical studies~\cite{BEC-BKT_PRL,VpairR_PRA}, the characteristic temperature for the BKT-BEC crossover was calculated by determining when the thermal excitation probability of a vortex pair in a sample becomes of order unity. Our results show qualitative agreement with the predictions, but their direct comparison is limited because the vortex detection efficiency in our experiments is not determined.

In conclusion, we have observed thermally activated vortex pairs in a 2D Bose gas trapped in a harmonic potential. This provides the clear confirmation of BKT superfluidity of the system. Ultracold atom experiments have been recently extended to 2D systems with Fermi gases~\cite{Fermi2d_Nature,Martin2D} or including disorder potentials~\cite{disorderBKT_PRA,disorderBKT_NJP}. The vortex detection method developed in this work will be an important tool to probe the microscopic properties of these systems. Integrated with an inteferometric technique~\cite{2DNIST_PRL,Roumpos11_NatPhys}, this method can be upgraded to be sensitive to the sign of a vortex.

We thank W.~J.~Kwon for experimental assistance. This work was supported by the NRF of Korea funded by MEST (Grants No. 2011-0017527, No. 2008-0062257, and No. WCU-R32-10045). We acknowledge support from the Global PhD Fellowship (J.C.), the Kwanjeong Scholarship (S.W.S.), and the T.J. Park Science Fellowship (Y.S.).


\vspace{6 in}

\begin{center}
\textbf{SUPPLEMENTAL MATERIAL}
\end{center}

\vspace{0.13in}

\noindent\textbf{Radial compression.} An additional radial confinement was provided by applying a magnetic quadrupole field to the trapped sample, where the atoms were in the $|F=1,m_F=-1\rangle$, weak-field seeking state. The symmetric axis of the magnetic quadrupole field was aligned with the center $z$-axis of the sample [Fig.~5(a)]. The trapping potential of the hybrid optical-magneto trap is described as
\begin{eqnarray}
\nonumber V(x,y,z) &=& \frac{1}{2}m (\omega_{x0}^2 x^2 + \omega_{y0}^2 y^2 + \omega_z^2 z^2) \\ \nonumber
&+& \frac{\mu_B}{2} B'_q\sqrt{\frac{x^2+y^2}{4} +  (z-z_0)^2} -mgz, 
\end{eqnarray}
where the first term corresponds to the optical harmonic potential, $\mu_B$ is the Bohr magneton, $B'_q$ is the magnetic field gradient along the axial direction, and $g$ is the gravitational acceleration. The position $z_0$ of the zero-field center was controlled with an external bias field $B_z$ along the $z$-direction as $z_0=B_z/B'_q$. The radial trapping frequencies are given as $\omega_{x,y}=\sqrt{\omega_{x0,y0}^2+ \frac{\mu_B {B'_q}^2}{8 m B_z}}$ at the trap center, and the aspect ratio of the trap $\lambda=\omega_z/\sqrt{\omega_x \omega_y}$. 

The sample preparation was carried out with $B'_q=0$ and $B_z=700$~mG ($\lambda=108$). We performed the radial compression by increasing the field gradient to $B'_q=7.8$~G/cm for 0.2~s and linearly ramping down the external bias field to $B_z=30$~mG for another 0.2~s [Fig.~5(b)]. The radial trapping frequencies of the compressed trap are $\omega_{x,y} \approx 2\pi\times$39~Hz. We observed that the atom number fraction of the coherent part reduces by about $7\%$ after this compression process. We note that the field gradient $B'_q$ needs to be kept smaller than 8.0~G/cm, corresponding to the strength to cancel the gravity, to prevent forming a local potential minimum at the zero-field point.
 
The experiments with a trap of $\lambda=63$ in Fig.~5(g)-(i) were carried out by preparing samples at $B'_{q}=2.0$~G/cm and $B_z\simeq 140$~mG, where the radial trapping frequencies at the trap center are estimated to be $(\omega_x, \omega_y) = 2\pi\times(5.6, 6.1)$~Hz.

\clearpage

\begin{figure*}[h]
\vspace{0.7in}
\begin{center}
\includegraphics[width=13cm]{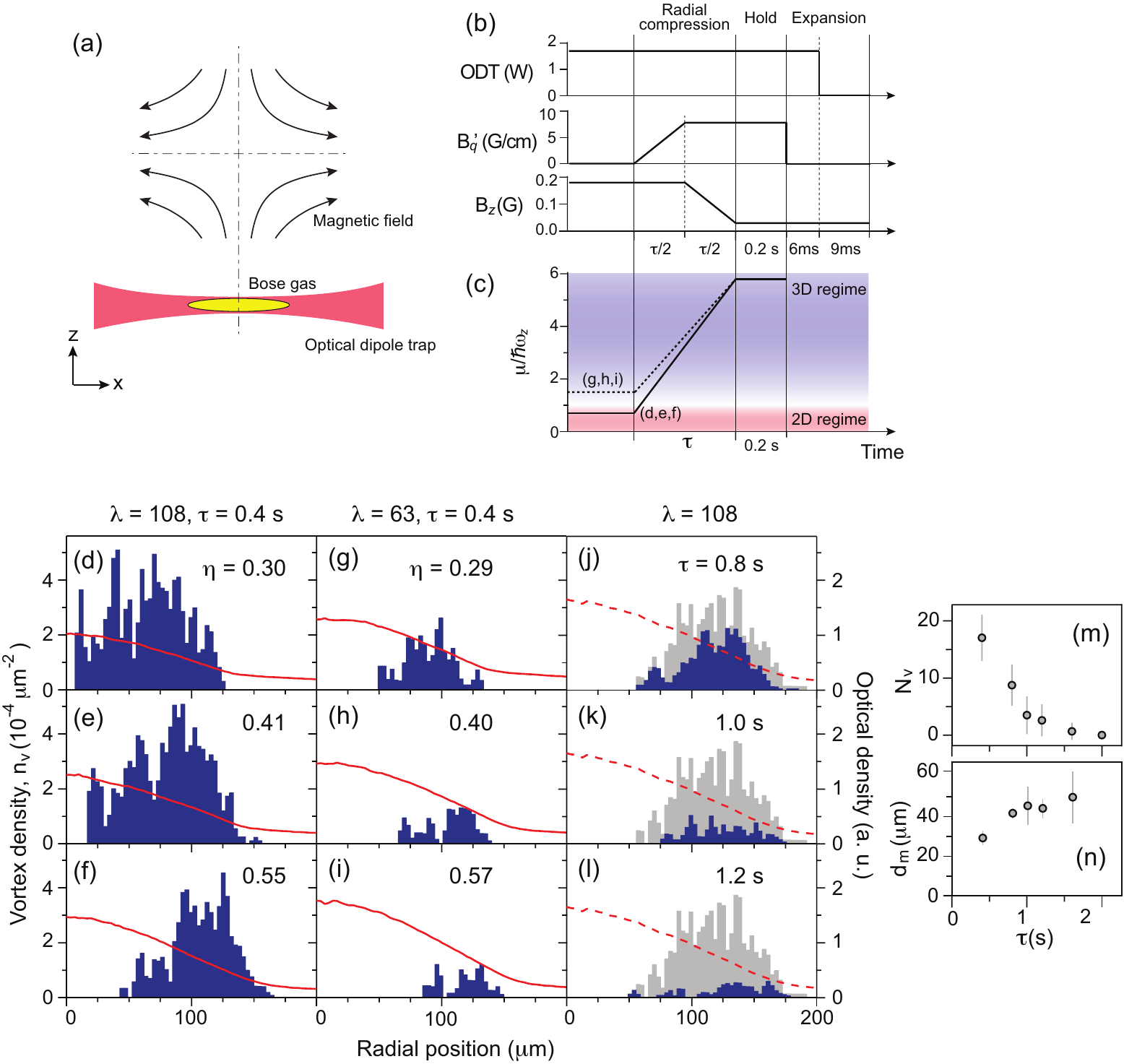}
\end{center}
\parbox{5.5in}{
\caption{(a) Schematic description of the experimental setup for the radial compression using a magnetic quadrupole field. (b) Time sequence of the compression and expansion procedure and (c) an illustration of the time evolution of the scaled chemical potential $\mu/\hbar\omega_z$ of the trapped sample. (d)-(l) The vortex density profiles and the atom density profiles for various sample conditions in terms of temperature ($\eta$ is the atom number fraction of the coherent part), trap aspect ratio $\lambda=\omega_z/\sqrt{\omega_x \omega_y}$, and the compression time $\tau$. The data in (d) and (f) are the same data in Fig.~3(d) and (e), respectively. The atom and vortex numbers are $(N, N_v)=(1.5(2)\times 10^6, 13(4))$ for (d), $(1.4(3)\times 10^6, 16(4))$ for (e),$(1.3(1)\times 10^6, 15(5))$ for (f), $(1.9(2)\times 10^6, 6(3))$ for (g), $(1.8(2)\times 10^6, 4(2))$ for (h), and $(1.7(2)\times 10^6, 3(2))$ for (i). The zero-temperature chemical potential of the samples in (g)-(i) is estimated to be $\mu\approx 1.5 \hbar \omega_z$. The data in (j)-(l) show the vortex density distributions for various compression times, obtained with samples with $N=1.8(2)\times 10^6$ and $\eta\approx0.55$. For comparison, the vortex density profile (gray) and the atom density profile (red) for $\tau=0.4$~s are displayed with other data. (m) The vortex number $N_v(\tau)$ and (n) the nearest vortex distance $d_m(\tau)$.}
} 
\label{sFigure1}
\end{figure*}

\begin{figure*}[h]
\vspace{1in}
\begin{center}
\includegraphics[width=8.5cm]{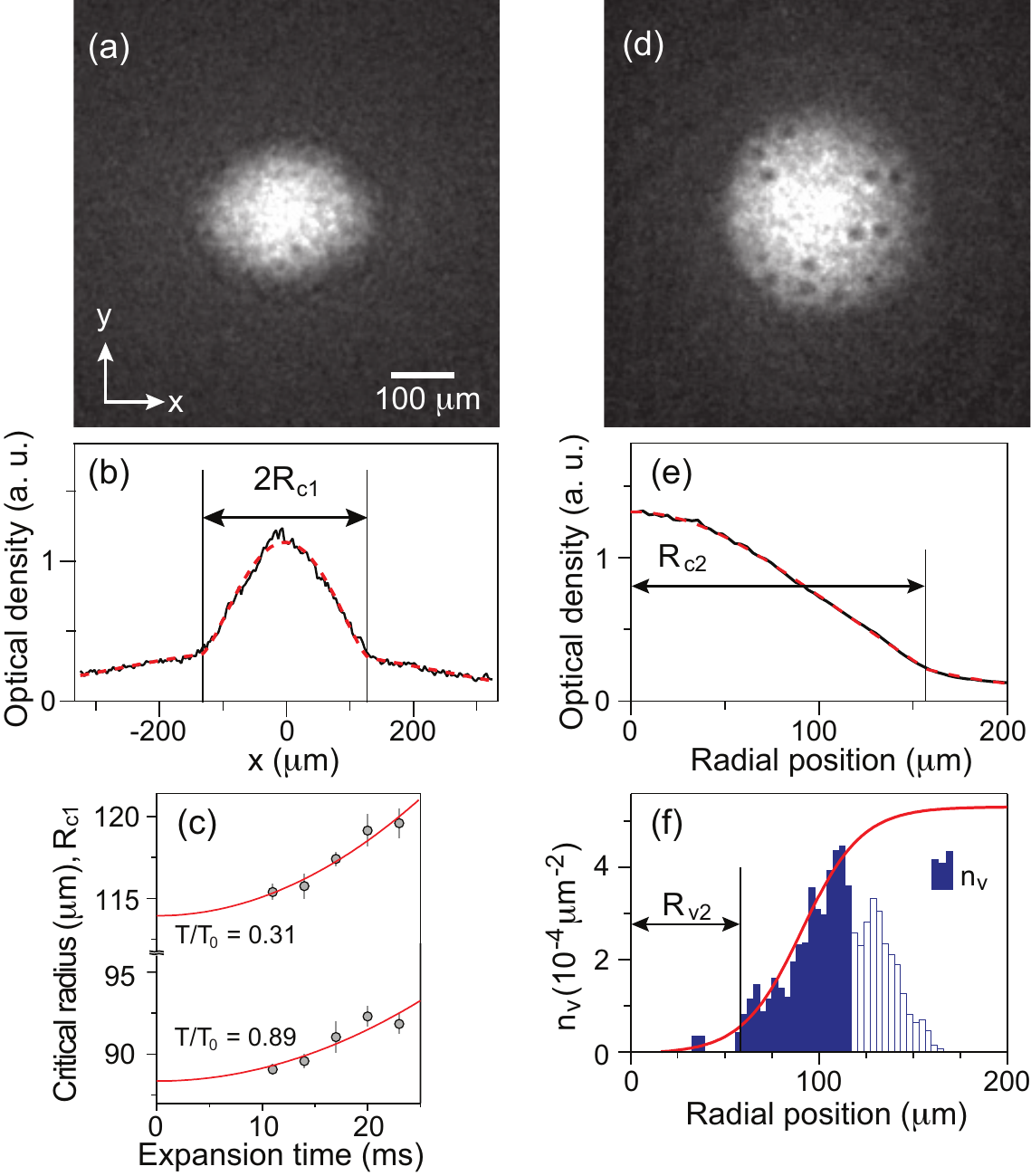}
\end{center}
\parbox{5in}{
\caption{(a) Optical density image of a quasi-2D sample after a 17-ms time-of-flight expansion [Fig.~1(a)] and (b) the density distribution integrated along the $y$-direction. The coherent part radius $R_{c1}$ is determined from a bimodal fit (red dashed line) to the density profile. (c) $R_{c1}(t)$ versus the expansion time $t$. The in-situ radius $R_{c0}$ of the coherent part is estimated from a fit (red line) of $R_c (1+\alpha t^2)$ to the data. (d) Image with applying the radial compression before expansion [Fig.~1(b)]. The expansion includes a 6-ms expansion in the optical trap and a subsequent 9-ms time-of-flight. (e) The coherent part radius $R_{c2}$ is determined from a bimodal fit (red dashed line) to the azimuthally averaged density profile and (f) the inner radius of the vortex region $R_{v2}$ is determined from a hyperbolic-tangent fit to the inner increasing part (blue solid) of the vortex density profile $n_v(r)$ with a threshold value of $6\times 10^{-5}~\mu$m$^{-2}$. The in-situ inner radius $R_{v0}$ of the vortex region is obtained as $R_{v0}=R_{c0} (R_{v2}/R_{c2})$, assuming that the relative position of the vortex region is preserved in the compression and expansion process. This assumption is supported by the observation in Fig.~5(j)-(l). The zero-temperature 2D Thomas-Fermi radius is calculated as $R_\text{TF}=(\frac{4\hbar^2N\tilde{g}\omega_y}{\pi m^2\omega_x^3})^{1/4}$ for the $x$-direction.}
}
\label{sFigure2}
\end{figure*}

\begin{figure*}[h]
\vspace{1in}
\begin{center}
\includegraphics[width=10cm]{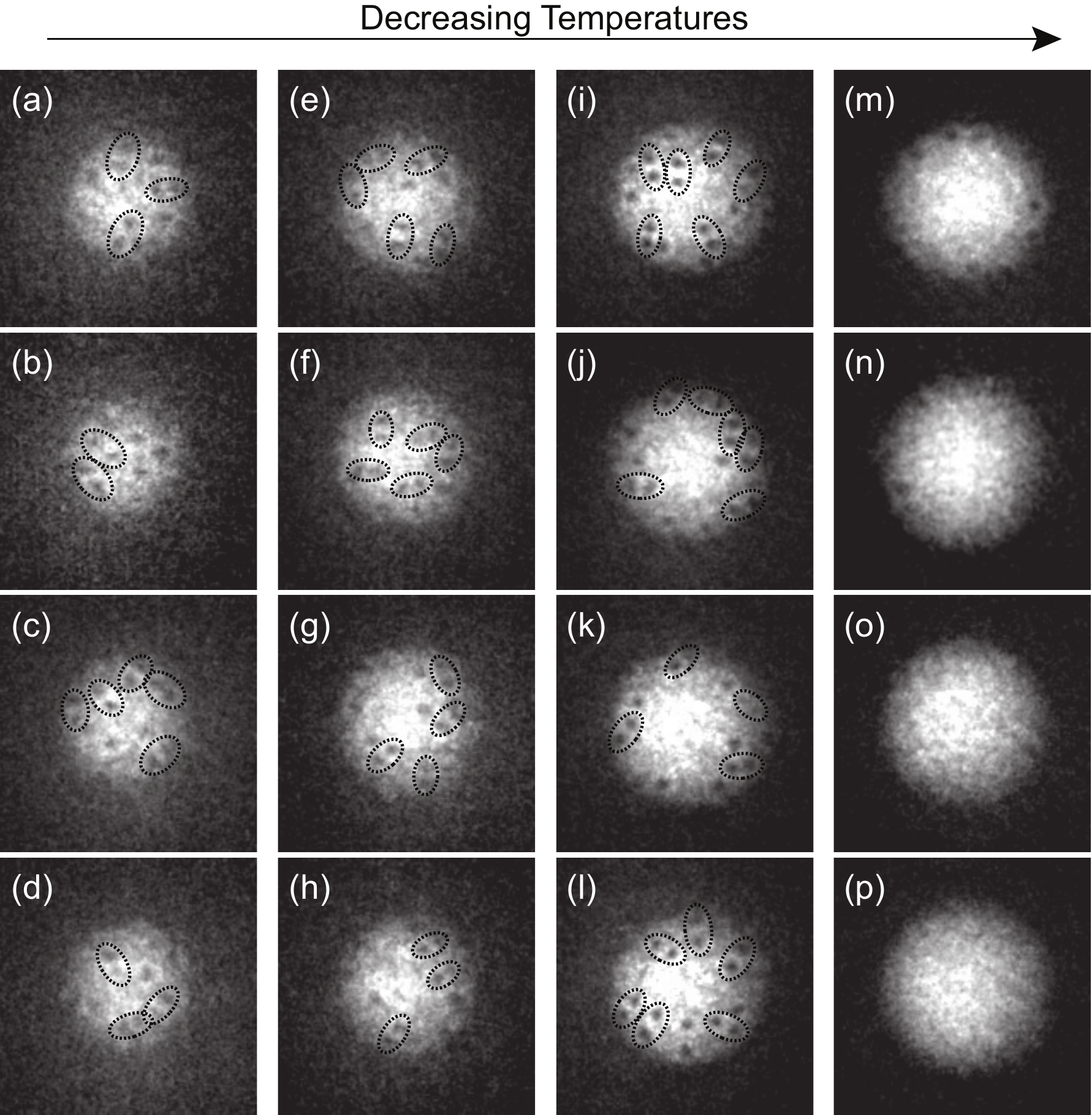}
\end{center}
\parbox{5in}{
\caption{Temperature dependence of the vortex distribution in a quasi-2D Bose gas trapped in a harmonic potential. Each column shows four images acquired for the same experimental sequence. The sample conditions are $(N,T)= (1.5(2)\times10^6, 67(12)$~nK) for (a)-(d), $(1.4(3)\times10^6, 56(6)~$nK) for (e)-(h), $(1.3(1)\times10^6, 48(8)$~nK) for (i)-(l), and $(1.0(3)\times10^6, 23(4)$~nK) for (m)-(p). The dotted lines indicate vortex pairs whose separation is less than $40~\mu$m.}
} 
\label{sFigure3}
\end{figure*}


\begin{references}
\bibitem{footnote1}
The BEC temperature $T_0$ is calculated from the relation $N=\frac{k^2 T_0^2}{\hbar^2 \omega_x \omega_y}\sum_{\nu=0}^{\infty} \textrm{Li}_2(e^{-\nu\frac{\hbar\omega_z}{kT_0}})$~\cite{Holzmann_EPL}, considering thermal population in the tight direction, where $\textrm{Li}_2(z)$ is the polylogarithm of order 2.
\end{references}
\end{document}